\documentclass[12pt]{iopart}
\usepackage{iopams}  
\usepackage{graphicx}        
\usepackage[numbers, sort&compress]{natbib}
\def\newblock{\hskip .11em plus .33em minus .07em}

\begin{document}

\title[Microscopic time-dependent approaches]{Challenges in description of heavy-ion collisions with microscopic time-dependent approaches}

\author{C\'edric Simenel}

\address{Department of Nuclear Physics, Research School of Physics and Engineering, Australian National University, Canberra, Australian Capital Territory 0200, Australia}
\ead{cedric.simenel@anu.edu.au}
\begin{abstract}
Important efforts have been dedicated in the past few years to describe near-barrier heavy-ion collisions with microscopic quantum theories like the time-dependent Hartree-Fock approach and some of its extensions. 
However, this field is still facing important challenges such as the description of cluster dynamics, the prediction of fragment characteristics in damped collisions, and sub-barrier fusion by quantum tunnelling. 
These challenges are discussed and possible approaches to solve them are presented.
\end{abstract}


\section{Introduction}

Nuclear structure studies involve microscopic approaches to describe many-nucleons in interactions (the nucleons are assumed to be structureless) \cite{rin80}.  
The time-dependent extension of such approaches to describe collisions between nuclei is one of the major challenges to nuclear theorists. 
In fact, the complexity of the nuclear quantum many-body problem has limited the time-dependent descriptions of nuclear dynamics essentially to mean-field approaches, with few exception such as the time-dependent generator coordinate method (TDGCM) \cite{ber84}, the time-dependent random-phase approximation (TDRPA) \cite{bal84}, the extended time-dependent Hartree-Fock (ETDHF) model \cite{won78,lac04} and the time-dependent density-matrix (TDDM) theory \cite{cas90,gon90}.
In analogy with the description of the dynamics of systems of electrons, one could consider the time-dependent density-functional theory (TDDFT) as well \cite{run84}. 
Indeed, the Hohenberg-Kohn theorem \cite{hoh64} has recently been extended to self-bound systems such as atomic nuclei \cite{mes09a,mes09b}.
In fact, mean-field models in nuclear physics are often based on energy density functional approaches which present strong similarities with TDDFT. 

The time-dependent Hartree-Fock (TDHF) theory is a self-consistent mean-field formalism initially developped by Dirac in 1930 to describe atoms \cite{dir30}.
Following the success of its static counterpart in nuclear structure \cite{vau72}, it has been applied to investigate nuclear dynamics since the mid-70s \cite{bon76}. 
However, despite several successes, early calculations \cite{neg82} suffered from computational limitations.

The increase of numerical power led to the recent development of three-dimensional TDHF codes \cite{kim97,mar05,uma05,nak05,ass09,seb09,eba10,ste11,has12,fra12,sca13,sek13}.
These codes have been used to investigate nuclear vibrations \cite{sim03,mar05,uma05,nak05,sim09,eba10,ste11,has12,fra12,sca13b} and heavy-ion collisions near the barrier such as the fusion process \cite{kim97,sim01,uma06a,mar06,sim07,guo12,sim13b}, transfer reactions \cite{uma08b,sim10b,sek13,sca13}, deep-inelastic collisions \cite{iwa10a,sim11}, clustering \cite{uma10a,leb12,sim13a}, and actinide collisions \cite{gol09,ked10}. 
In many cases, the predictive power of modern TDHF calculations is very good (see \cite{sim12b} for a recent review). 
Some codes have also been used to investigate the dynamics of neutron-star crusts \cite{seb09,sch13}.

One important advantage of fully microscopic theories like TDHF is that their only input is the set of parameters of the energy density functional describing the interaction between nucleons, such as the Skyrme functional \cite{sky56}. 
As a consequence, the calculations do not rely on measured quantities which are specific to the studied system, such as excited states of the collision partners or their nucleus-nucleus potential.
This point is crucial for reactions involving exotic nuclei for which little is known. 

Despite successes,  mean-field descriptions present  strong limitations.
One of them is the restriction to independent (quasi-)particle states.
Clustering effects (e.g., alpha-cluster configurations), which are often essential in light and/or weakly bound nuclei are then usually strongly underestimated. 
As a result, cluster break-up and cluster transfer reactions are poorly described at the mean-field level. 
Another limitation is the fact that the TDHF theory is optimised for expectation values of one-body observables only \cite{bal81}. 
Indeed, in some cases, fluctuations of such observables are highly underestimated \cite{koo77}.
Last but not least, in its standard real time formulation, the TDHF theory is unable to describe the tunnelling of the many-body wave-function. 
As a consequence, there is no sub-barrier fusion, which is probably the main drawback of this approach. 

These selected limitations are discussed in the following sections. 
For each of these problems, the present status and challenges are presented in more details.
Possible approaches to overcome these limitations are also discussed.

\section{Cluster dynamics}

It is well known that the residual interaction between nucleons can generate cluster structures in nuclei. 
The simplest cluster is a pair of nucleons. 
It can be generated by the pairing residual interaction. 
The TDHF theory is based on an independent particle picture in which pairing effects are neglected.
However, pairing correlations can be included at the mean-field level by considering quasi-particle vaccua instead of independent particle states in the variational space \cite{rin80}.

Pairing correlations have then been included in extensions of TDHF codes at the BCS (Bardeen-Cooper-Schrieffer) level \cite{eba10,sca13}
and using the more general TDHF-Bogoliubov (TDHFB) theory \cite{ave08,ste11,has12}. 
These correlations are particularly important in the description of multi-nucleon transfer at near and sub-barrier energies \cite{cor09,oer01}.
As a result, time-dependent description of pairing correlations have successfully reproduced the enhancement of pair-transfer \cite{sca13}, which are usually strongly underestimated at the TDHF level \cite{sim10b,eve11}.

Other types of clusters, such as alpha particles, could also affect the reaction mechanisms.
Note that these effects are often magnified in the case of exotic nuclei for which the clusters are usually more weakly bound.
In addition to multi-nucleon transfer, the latter can also more easily break-up when interacting with a collision partner. 

The dynamics of clusters can be studied at the mean-field level only when these clusters are present in the initial wave-function of the system (see, e.g., \cite{uma10a} for a study of three alpha-clusters dynamics).
However, in most cases, there is no clustering in the mean-field states describing the ground-states of the collision partners. 
Thus, the study of cluster dynamics in reactions implies to extend the present mean-field formalism.

One possible approach to overcome this limitation would be to consider a time-dependent extension of the modern version of the fermionic molecular dynamics (FMD) model \cite{rot04}.
In the FMD approach, single-particle states are constrained to be Gaussian wave-packets. 
This limitation of the variational space allows the use of advanced beyond mean-field techniques such as  angular momentum projection and generator coordinate method (GCM) \cite{hil53} to describe the structure of the nuclei. 
Alpha-clustering appears naturally in this approach, which has been successful in describing excited states such as the Hoyle state in $^{12}$C \cite{che07}. 

It is tempting to envisage a time-dependent extension of this static FMD beyond mean-field model in order to describe the dynamics of clusters in heavy-ion collisions. 
In particular, one could investigate alpha-transfer and break-up mechanisms. 
However, developing such a time-dependent extension is not without difficulty. 
Indeed, the state of the system is represented by a superposition of Slater determinants.
Each of these determinants has a weight determined by the projection and GCM calculations. 
The difficulty is that both the Slater determinants and their associated weights are expected to evolve in time. 
A possible approach would be to consider a mean-field like evolution of each determinant and to solve the time-dependence of their weights with the TDGCM\footnote{This approach will be discussed in more details in the next section in the case of Slater determinants evolving according to the TDHF equation.}.

\section{Characteristics of fragments in damped collisions}

As in the cluster transfer and break-up reactions discussed above, experimentally, one only has access to the final products of the reaction. 
Important quantities which are used to characterise the reaction products are the number of fragments, their charge, mass, kinetic energy and angular momentum. 
These quantities are essentially associated with one-body observables $\hat{Q}$. 
Ideally, one would like to be able to predict the distribution of probabilities $p_i$ associated to the eigenvalue $q_i$ of $\hat{Q}$.

Balian and V\'en\'eroni have shown that the TDHF(B) theory is a mean-field approach optimised to the determination of the expected values of one-body observables $\langle\hat{Q}\rangle$ \cite{bal81}, that is, the centroid of the distribution of $q_i$. 
As a consequence, mean-field theories are not optimised to the prediction of the fluctuations of $\hat{Q}$.
Indeed, such fluctuations are quantified by the standard deviation of the distribution of $q_i$,
\begin{eqnarray} 
\sigma_Q&=&\sqrt{\sum_ip_iq_i^2-\left(\sum_ip_iq_i\right)^2}=\sqrt{\langle\hat{Q}^2\rangle-\langle\hat{Q}\rangle^2}.\label{eq:1}
\end{eqnarray}
As we can see, it does not only involve the expectation value of a one-body operator $\langle\hat{Q}\rangle$, but also the expectation value of its {\it square} $\langle\hat{Q}^2\rangle$.

To overcome this limitation, Balian and V\'en\'eroni have derived an equation for $\sigma_Q$ which is equivalent to the TDRPA \cite{bal84}. 
In this approximation, small fluctuations of the observable of interest are included around the mean-field evolution. 
Recent applications to $^{40}$Ca+$^{40}$Ca deep-inelastic collisions have led to a good agreement with experimental data \cite{sim11}. 
Note also that fluctuations in the initial state can be included in a stochastic mean-field approach \cite{ayi08,lac13}, which reduces to the TDRPA in the small amplitude limit \cite{ayi08}. 
This approach has been recently applied to heavy-ion collisions in the semi-classical limit \cite{was09b,ayi10,yil11}.

Nevertheless, the TDRPA is not able to describe the entire distribution of probabilities. 
For example, it cannot predict an eventual skewness  or bimodality of the distribution. 
The prediction of such distributions remains an important challenge in the field. 

\begin{figure}[htbp] 
   \centering
   \includegraphics[width=5in]{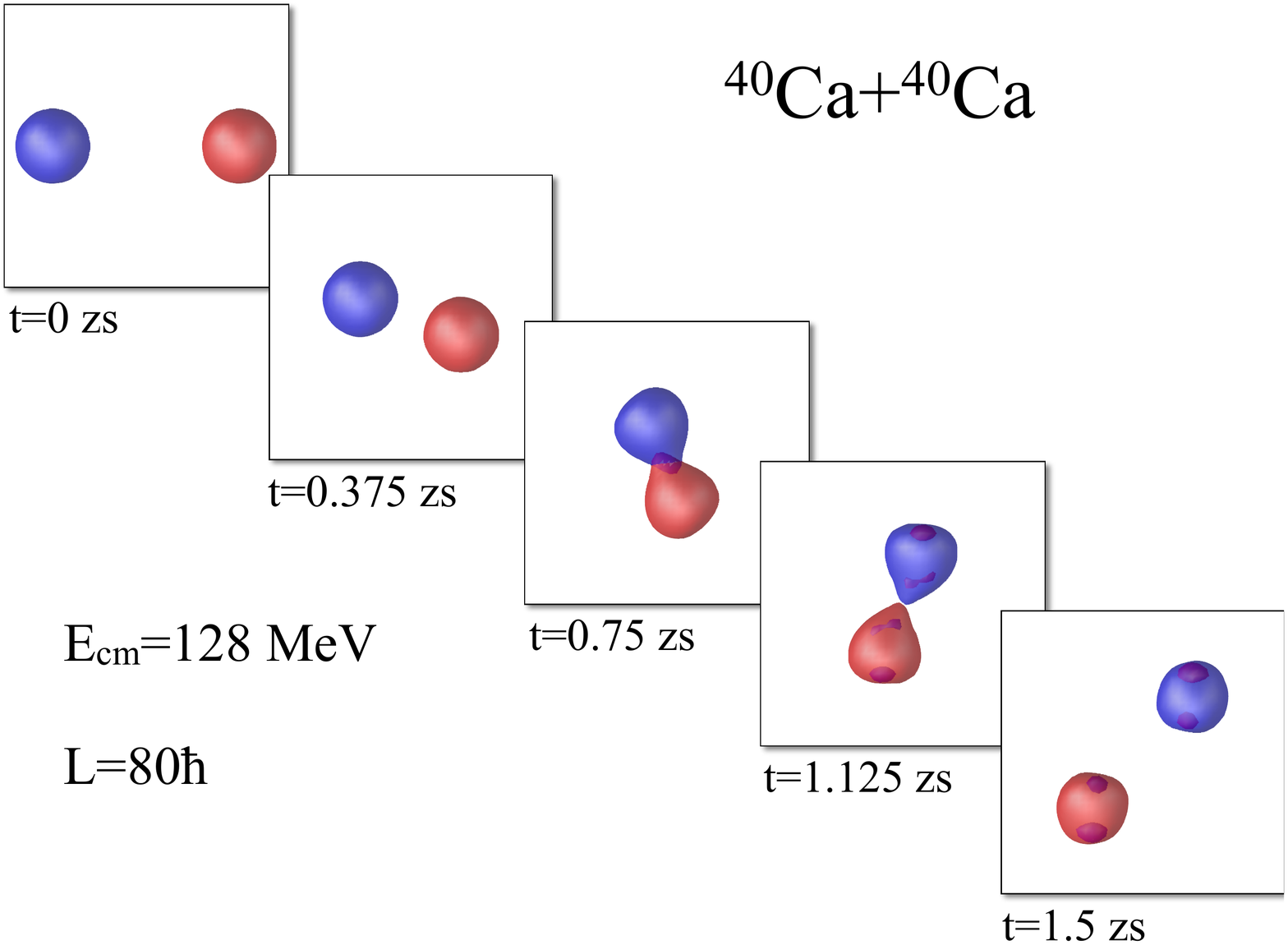} 
   \caption{Isodensities at $\rho=0.03$~fm$^{-3}$ for the $^{40}$Ca$+^{40}$Ca collision at $E_{cm}=128$~MeV and $L=80\hbar$ calculated with the TDHF approach.}
   \label{fig:Ca+Ca_TDHF}
\end{figure}

Let us take the example of transfer reactions.
To some extent, the latter can be described at the TDHF level.
An example is shown in figure~\ref{fig:Ca+Ca_TDHF} for a $^{40}$Ca$+^{40}$Ca collision at $E_{cm}=128$~MeV (approximatively 2.5 times the barrier height) and $L=80\hbar$.
The final state of the system is a coherent superposition of different transfer channels. 
The main drawback of this approach is that all channels are described by the same mean-field.
In particular, this mean-field is optimised for the evolution of fragments having the average mass and charge of the final distributions. 
It is clear that transfer channels associated with fragment masses and charges deviating significantly from the average values are expected to be poorly described in this approach.
This problem, known as {\it cross channel coupling}, was already identified in  early applications of the TDHF theory in nuclear physics \cite{koo77}.

Once again the TDGCM could provide an elegant solution to this problem. 
Instead of having one mean-field describing all channels, we assume that each transfer channel is obtained from its own mean-field evolution, that is, there is one mean-field per channel\footnote{An adiabatic version of the TDGCM has been applied in the past to study nuclear fission \cite{gou05}. In this case the independent quasi-particle states are obtained with static HFB calculations under an external constraint. To study nuclear reactions, however, it seems more natural to consider a mixing of time-dependent mean-field states. }. 
The total wave function could then be written as 
\begin{equation}
|\Psi(t)\rangle=\sum_iC_i(t)|\Phi_i(t)\rangle,
\label{eq:Psi}
\end{equation}
where $|\Phi_i(t)\rangle$ is an independent particle state associated to the transfer channel $i$. 
$|\Phi_i(t)\rangle$ could be obtained from the TDHF equation with an external potential {\it forcing} the transfer of $\delta n_i$ nucleons. 

For illustrational purpose, let us use the one-dimensional model introduced in \cite{bon76} to describe the collision of infinite slabs of nuclear matter.
Consider a symmetric collision along $x$ with a centre of mass at $x=0$.
The operator counting the number of particles in the right side ($x>0$) can be written
\begin{equation}
\hat{N}_R=\int dx \Theta(x) \hat{a}^\dagger(x)\hat{a}(x),
\end{equation}
where $\Theta(x)$ is a step function. 
The external potential can  be chosen as $\lambda_i\hat{N}_R$ where the Lagrange parameter $\lambda_i$ plays the role of a difference of chemical potentials between the two fragments.
The evolution of $|\Phi_i(t)\rangle$ is then obtained from the mean-field equation
\begin{equation}
i\hbar\frac{d}{dt}|\Phi_i\rangle=\left(\hat{H}_{MF}[\Phi_i]+\lambda_i\hat{N}_R\right)|\Phi_i\rangle,
\label{eq:MF}
\end{equation}
where 
$\hat{H}_{MF}$
is the mean-field Hamiltonian. 
The parameter $\lambda_i$ is adjusted to obtain the desired asymptotic expectation value of the particle number in the right fragment $\langle\hat{N}_R\rangle(t\rightarrow\infty)=n_i$. 
Note that some conservation laws such as translational invariance are broken by $\hat{N}_R$.
In particular, this could induce a spurious centre of mass motion which should be corrected for.

\begin{figure}[htbp] 
   \centering
   \includegraphics[width=3in]{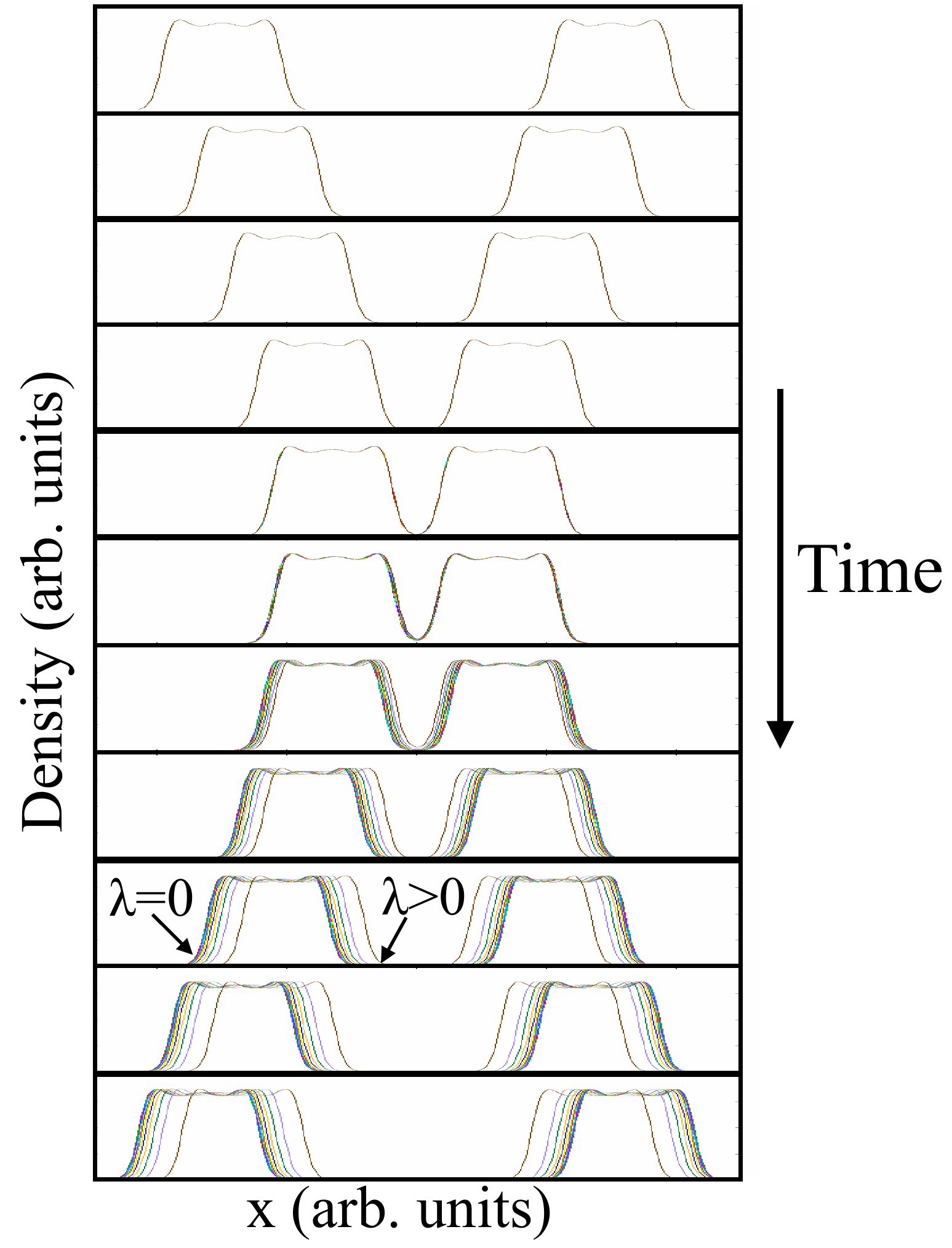} 
   \caption{Symmetric collision of nuclear slabs. The different density profiles are obtained by varying the strength $\lambda$ of the external potential.}
   \label{fig:Slabs}
\end{figure}

A numerical application is presented in figure~\ref{fig:Slabs}.
A long-range repulsive interaction is included on top of the short range nuclear interaction.
The collision occurs at an energy just below the barrier potential generated by the competition between these two interactions. 
The different density profiles are obtained by varying $\lambda_i$ which drives the amount of transfer from one slab to the other.
In the exit channel, the velocity of the fragments depends on the amount of transfer.
As a result, the spacial overlap between densities obtained with different $\lambda_i$ vanishes as time goes on. 
This effect is obviously absent if only one mean-field is used to describe all transfer channels as in TDHF. 

Once the $|\Phi_i(t)\rangle$ have been determined, the next step is to compute the weights $C_i(t)$ using the TDGCM. 
This is based on a variational principle requesting the stationarity of the action
\begin{equation}
S[\Psi;t_0,t_1]=\int_{t_0}^{t_1} dt \langle\Psi(t)|i\hbar\frac{d}{dt}-\hat{H}|\Psi(t)\rangle.
\end{equation}
We then seek for solutions $\Psi(t)$ obeying $\delta S=0$. 
Using equations (\ref{eq:Psi}) and (\ref{eq:MF}), the action can be rewritten as 
\numparts
\begin{eqnarray}
S[\Psi;t_0,t_1]&=\sum_{ij} \int_{t_0}^{t_1} dt C_i^*&\left[i\hbar  \dot{C}_j \langle\Phi_i|\Phi_j\rangle\right.\label{eq:S1}\\
&&+C_j\langle\Phi_i|(\hat{H}_{MF}+\lambda_j\hat{N}_R)|\Phi_j\rangle\label{eq:S2}\\
&&\left.-C_j\langle\Phi_i|\hat{H}|\Phi_j\rangle\right].\label{eq:S3}
\end{eqnarray}
\endnumparts

\begin{figure}[htbp] 
   \centering
   \includegraphics[width=3in]{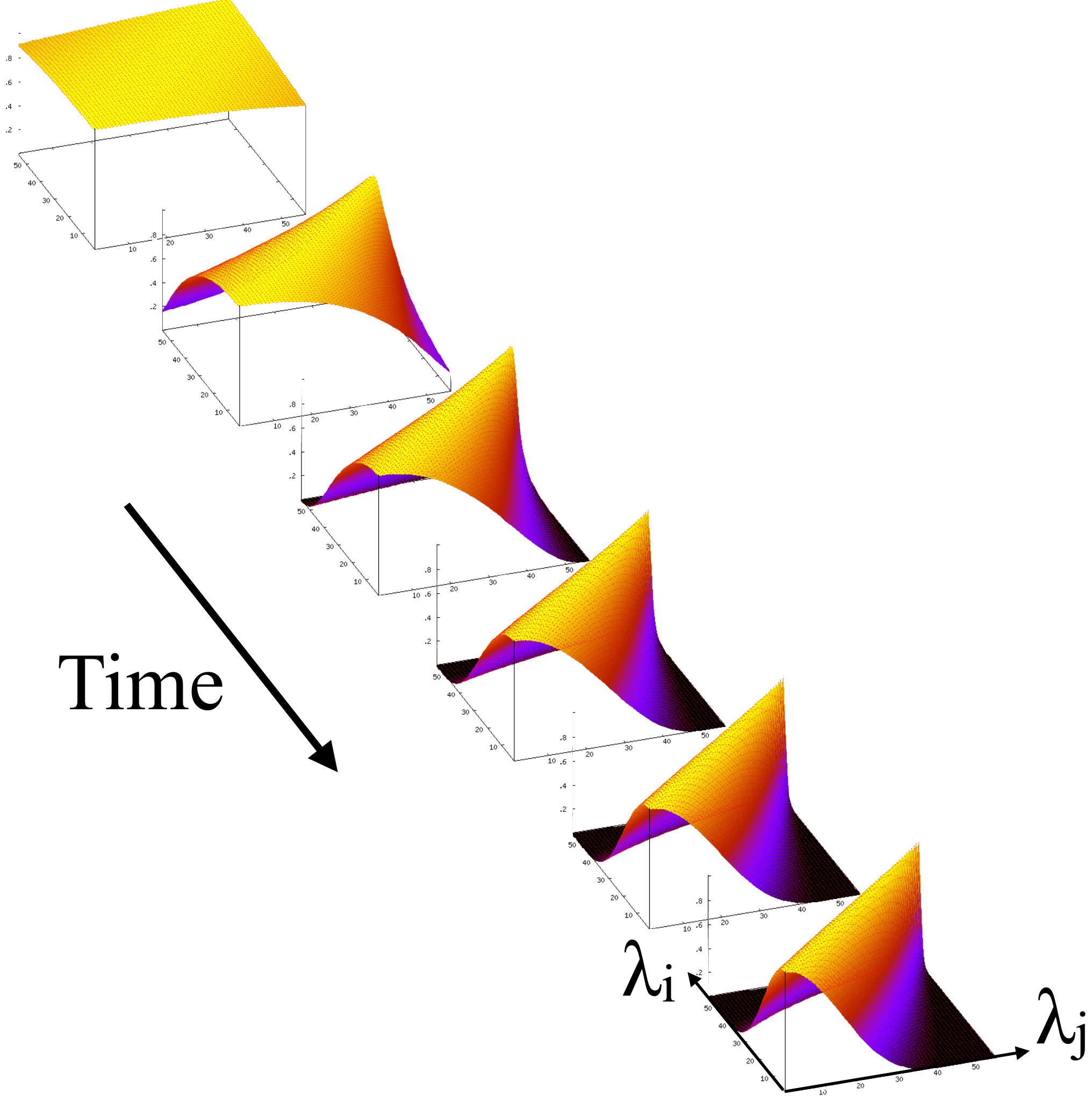} 
   \caption{Absolute value of the matrix elements  $\langle\Phi_i|\Phi_j\rangle$ (vertical axis) associated to the six last density profiles of the slab collision shown in figure~\ref{fig:Slabs}. }
   \label{fig:overlap}
\end{figure}

Solving the variational principle then requires the overlap matrices to be determined.
The two first, in equations~(\ref{eq:S1}) and~(\ref{eq:S2}), do not present any difficulty as $\hat{N}_R$ and $\hat{H}_{MF}$ are one-body operators. 
For illustration, the absolute value of the matrix elements  $\langle\Phi_i|\Phi_j\rangle$ is shown in figure~\ref{fig:overlap} for the previous example of slabs collision. 
The off-diagonal terms vanish rapidly after the collision due to the spatial separation of the outgoing slabs for different $\lambda$.  
However, the matrix elements in equation~(\ref{eq:S3}) are much more complicated to compute as $\hat{H}$ is in general a many-body operator. 
Moreover, one should be careful when computing the off-diagonal elements of $\hat{H}$. 
Indeed, density-dependent  effective interactions such as the Skyrme one can lead to spurious effects in beyond mean-field calculations, where matrix elements of the effective Hamiltonian between different Slater determinants (or quasi-particle vaccua) need to be computed \cite{dob07,ben09,dug09}. 
Possible solutions have been proposed to regularise the problem and allow for the use of standard energy density functionals \cite{lac09}. 
Alternatively, one could construct the functional as an expectation value of a strict Hamiltonian, in particular without density-dependent terms \cite{sad13}. 

\section{Sub-barrier fusion}

The above discussion showed that constraining all channels to evolve with the same mean-field, as in the TDHF approach, is a strong limitation.
The situation is even worse in the case of sub-barrier fusion reactions. 
Before discussing this problem, let us first present briefly the present status of research in low-energy fusion.

In sub-barrier collisions, most of the flux goes into quasi-elastic reaction channels. 
However, due to quantum tunnelling, there is a non-zero probability for fusion to occur at energies below the Coulomb barrier.
Note that, in many cases, the system has more than one barrier \cite{das98} due to the couplings between relative motion and internal degrees of freedom \cite{das85}.
Near barrier fusion is then usually treated using a macroscopic coupled-channels approach \cite{hag12}.
This approach requires the knowledge of the  structure of the collision partners as well as their interaction potential.
One problem is that these quantities are not always known, in particular for reactions involving exotic nuclei. 
A possible solution of this problem is to compute these parameters directly with TDHF \cite{uma06b,was08,sim13c} and use them in standard coupled channel calculations \cite{sim13c}.

However, some difficulties remain. 
For instance, recent observations of deep sub-barrier fusion hindrance (as compared to standard coupled-channels calculations) \cite{jia04,das07} have led to questioning our understanding of quantum tunnelling in fusion of heavy nuclei \cite{das03,mis06,das07,ich09b}.
It is thus highly desirable to achieve a fully microscopic description of quantum many-body tunnelling.

\begin{figure}[htbp] 
   \centering
   \includegraphics[width=2in]{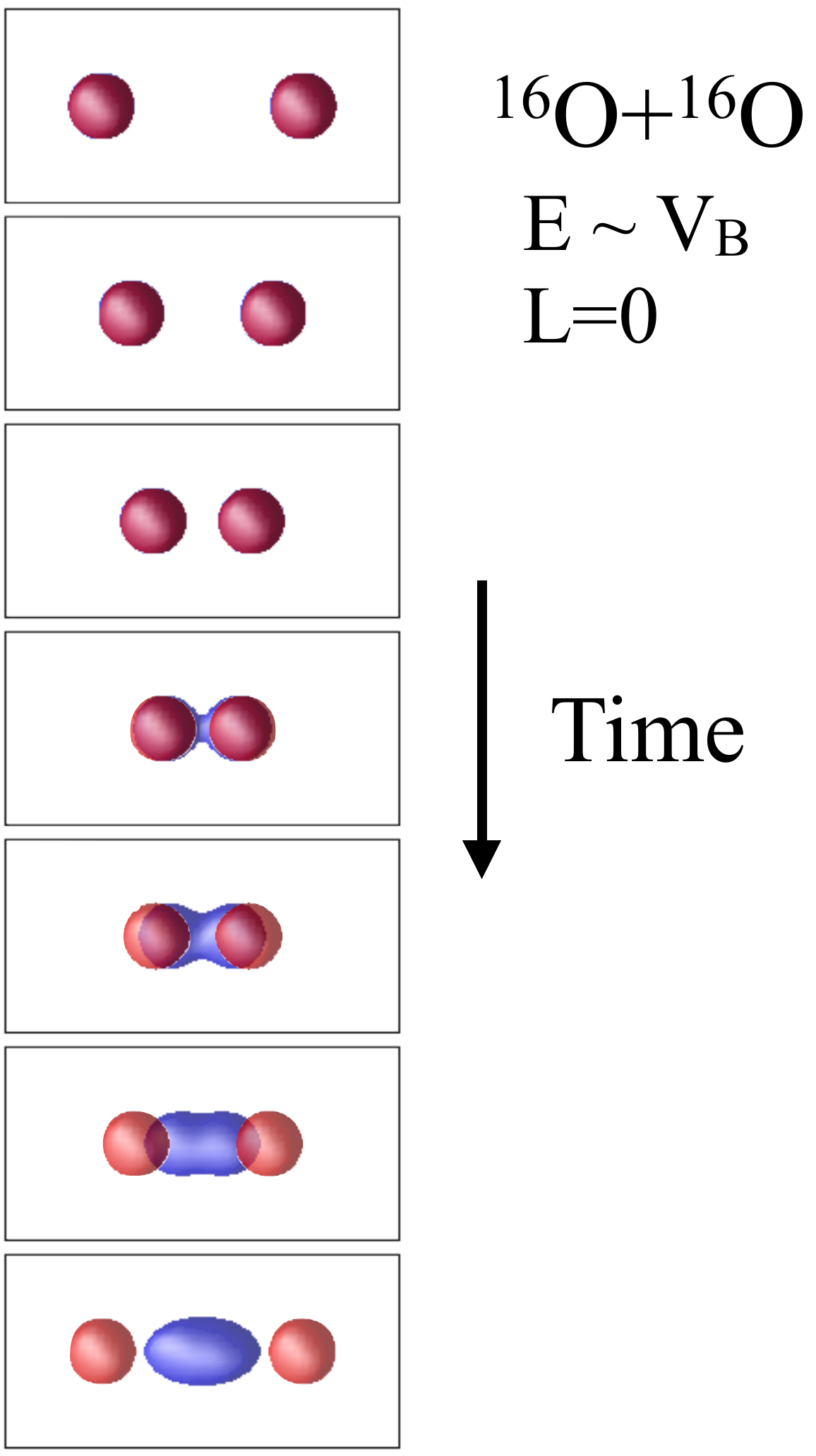} 
   \caption{Isodensities  (one just above the barrier, leading to fusion, and one just below, leading to reseparation of the fragments) for  $^{16}$O+$^{16}$O central collisions computed with TDHF.}
   \label{fig:fusion}
\end{figure}

This brings us back to our problem.
Due to its mean-field nature, the TDHF theory is unable to describe the tunnelling of the many-body wave function. 
As a consequence, in a single TDHF calculation of a heavy-ion collision, the fusion probability is either 0 or 1. 
To get intermediate values, one needs in principle, an approach with at least two Slater determinants: one leading to fusion, and one to the reseparation of the fragments. 
This is illustrated in figure~\ref{fig:fusion}, where two TDHF density evolutions are shown for  $^{16}$O+$^{16}$O central collisions.
Just above the barrier the system fuses, while just below it reseparates in two fragments. 
In reality, the system should be in a coherent superposition of these two mean-field states.

Naturally, one would think the TDGCM could be used to solve this problem\footnote{An adiabatic version of the GCM has also been considered in the case of fusion of light nuclei at low energy \cite{ber80,des89}.}. 
As described in the previous section, in this approach the many-body wave function is in a superposition of mean-field states evolving according to their own TDHF trajectory.
If some of these trajectories lead to fusion, while the others do not, then the fusion probability is between 0 and 1. 
The main open question in this approach is the choice of the collective variable.
The latter is crucial as it determines the external potential leading to a differentiation of the mean-field evolutions. 

An alternative choice to the TDGCM to describe tunnelling microscopically has been proposed in the early 80's \cite{lev80c,rei81b}. 
It is based on Feynman's path integral approach to quantum mechanics \cite{fey48}.
This can be applied to many-body systems and, in the stationary phase approximation (SPA) time-dependent mean-field equations are recovered \cite{lev80a}.
It is interesting to note that, for a single-particle system, the SPA leads to classical mechanics.
The fact that the TDHF approach does not include quantum many-body tunnelling is then a classical behaviour induced by the use of the SPA. 

It is well known that a semi-classical approximation to quantum tunnelling of a single-particle across a barrier potential, similar to the WKB formula, can be derived from the Feynman path integral formalism in {\it imaginary time} with the SPA (see, e.g., \cite{neg82}).
By analogy, one can describe quantum tunnelling of a many-body wave function at the {\it mean field} level using the SPA and imaginary time propagation \cite{lev80c,rei81b}. 
The resulting equations are much more complicated than the usual real time mean-field equations.
Indeed, they consist of a set of coupled integro-differential equations in both space and time. 
As a proof of principle, few schematic applications have been performed in the case of spontaneous fission \cite{lev80c,neg82,neg89}.
However, practical applications have been limited by the difficult task of finding many-body closed trajectories in imaginary time.

Of course, similar difficulties are expected in the application of this method to sub-barrier fusion reactions. 
Moreover, the transition between real-time and imaginary time propagation is an additional problem. 
Indeed, the initial configuration is two nuclei moving toward each other (see top panels of figure~\ref{fig:fusion}), which can be treated in real time, while the tunnelling through the barrier involves imaginary time propagation. 
As the nucleus-nucleus potential is not uniquely defined (and one would like to avoid introducing a macroscopic variable which is needed to define such a potential), the transition between classically allowed and classically forbidden regions is somewhat arbitrary. 
This is then also true for the transition between real and imaginary time propagation, 
To solve this problem, one would need to consider more general mean-field equations involving complex-time propagation instead of purely real or purely imaginary time evolutions.

\section{Conclusions}

Some challenges in the description of nuclear reactions with microscopic approaches have been presented. 
Clustering effects are present in the structure of some light and/or exotic nuclei.
The dynamics of such clusters, such as alpha-particles, is essential in transfer and break-up reactions. 
Unfortunately, clustering effects are usually not included in standard mean-field approaches such as TDHF. 
Beyond Mean-field models for nuclear structure, such as a modern version of the FMD, could be extended to incorporate time dependence in order to simulate cluster dynamics in collisions with the TDGCM. 
The TDGCM could also be used to describe transfer channels in their own mean-field (instead of one mean-field for all channels as it is the case in the TDHF theory) and, then improve the descriptions of fragment characteristics in damped collisions. 
In the case of sub-barrier fusion, which cannot be studied with TDHF due to a lack of many-body tunnelling, an alternative approach to the TDGCM is to consider Feynman path integrals for the many-body system in complex time with the stationary phase approximation. 

All these approaches face both technical and conceptual difficulties, such as the choice of collective coordinates in the TDGCM, and the transition from real to imaginary time evolution in the path integral formulation.
The development of high performance computing facilities will certainly help in performing realistic applications.
The latter will be a great asset to the experimental programs with low-energy rare isotope beam facilities. 

\section*{Acknowledgements}
The author is grateful to D. J. Hinde and M. Dasgupta for useful discussions and for their support. 
S. Umar is thanked for his comments on the manuscript. 
H. Smith is also warmly thanked for her in depth work on Feynman path integrals. 
This work has been supported by the Australian Research Council by the Future Fellowship FT120100760, Laureate Fellowship FL110100098 and Discovery grants DP1094947 and DP110102879. 
The calculations have been performed on the NCI National Facility in Canberra, Australia, which is supported by the Australian Commonwealth Government.


\end{document}